\documentclass[%
 reprint,
 bibnotes,
 amsmath,amssymb,
 aps,
 prl,
]{revtex4-1}

\usepackage{graphicx}
\usepackage{dcolumn}
\usepackage{bm}

\usepackage{graphicx}
\usepackage{epstopdf}
\usepackage{subfigure}
\usepackage{upgreek}
\usepackage{natbib}

\begin{document}
\def\pmb#1{\setbox0=\hbox{#1}%
\kern-.025em\copy0\kern-\wd0 \kern.05em\copy0\kern-\wd0
\kern-.025em\raise.0433em\box0 }
\def\pdd#1#2{\frac{\partial #1}{\partial #2}}
\def\p{\par}
\def\nd{\noindent}

\def\cP{{\cal{P}}}

\def\drabp{\delta \pmb{$r$}^{\prime}_{\alpha \beta}}
\def\drap{\delta \pmb{$r$}^{\prime}_{\alpha}}

\def\Fab{\pmb{$F$}_{\alpha \beta}}

\def\npot{{n + {1 \over 2}}}
\def\noot{{n + {1 \over 2}}}
\def\npo{{n + 1}}

\def\plp{\pmb{$l$}_\parallel}
\def\plv{\pmb{$l$}_\perp}
\def\wtplv{\widetilde{\pl}_\perp}
\def\wtplp{\widetilde{\pl}_\parallel}
\def\pfp{\pmb{$f$}_\parallel}
\def\pfv{\pmb{$f$}_\perp}
\def\pld{\pmb{$l$}'}
\def\pldv{\pmb{$l$}'_\perp}
\def\pldp{\pmb{$l$}'_\parallel}
\def\pLv{\pmb{$L$}_\perp}
\def\pLp{\pmb{$L$}_\parallel}

\def\sx{\pmb{{\cmmi x}}}
\def\sc{\pmb{{\cmmi c}}}
\def\sk{\pmb{{\cmmi k}}}
\def\sr{\pmb{{\cmmi r}}}
\def\sth{\pmb{{\cmmi $\theta$}}}
\def\sK{\pmb{{\cmmi K}}}

\def\vpi{\varphi}

\def \vee {\vec e}
\def \vW {\vec W}
\def \vn {\vec n}
\def \vW {\vec W}
\def \vf {\vec f}
\def \vF {\vec F}
\def \vex {\vec x}
\def \vom {\vec \omega}
\def \vsi {\vec \sigma}
\def \vov {(\vec \omega, \vartheta)}
\def \vA{\vec A}
\def \vv{\vec v}
\def \vt {\vec \tau}
\def \va{\vec a}
\def \wtd{\widetilde {\bf D}}
\def \vtt{\vec t}
\def \vn {\vec n}
\def \vF {\vec F}
\def \vex {\vec x}
\def \vom {\vec \omega}
\def \vsi {\vec \sigma}
\def \vov {(\vec \omega, \vartheta)}
\def \vA{\vec A}
\def \vv{\vec v}
\def \vt {\vec \tau}
\def \va{\vec a}
\def \vv{\vec v}

\def\ptth{\widehat{\pmb{$t$}}}
\def\wtpldv{\widetilde{\pl}'_\perp}
\def\wtpldp{\widetilde{\pl}'_\parallel}

\def\wtL{\widetilde{L}}
\def\wtpLv{\widetilde{\pL}_\perp}
\def\wtpLp{\widetilde{\pL}_\parallel}
\def\wtu{\widetilde{u}}
\def\wtv{\widetilde{v}}
\def\wtw{\widetilde{w}}
\def \wtA{\widetilde {A}}
\def \wtB{\widetilde {B}}
\def \wtC{\widetilde {C}}
\def \wtD{\widetilde {D}}
\def\wtU{\widetilde{U}}
\def\wtV{\widetilde{V}}
\def\wtW{\widetilde{W}}
\def\wtX{\widetilde{X}}
\def\wtY{\widetilde{Y}}
\def\wtZ{\widetilde{Z}}
\def\wtOm{\widetilde{\Omega}}
\def\wtom{\widetilde{\omega}}
\def\wtxi{\widetilde{\xi}}
\def\wtP{\widetilde{P}}
\def\wpx{\widetilde{\px}}
\def\wpal{\widetilde{\pal}}
\def\wpo{\widetilde{\po}}
\def\wpL{\widetilde{\pL}}
\def\wpu{\widetilde{\pu}}
\def\wg{\widetilde{G}}
\def\wgs{\widetilde{G}^*}
\def\gs{G^{*}}

\def\wu{\widehat u}
\def\wy{\widehat y}
\def\wx{\widehat x}
\def\wv{\widehat v}

\def\Dt{\Delta t}

\def\Fa{F_{\alpha}}
\def\oF{\overline{F}}
\def\fap{f^{\prime}_{\alpha}}
\def\ofp{{\overline{f}}^{\prime}}
\def\Ua{\pmb{$U$}_{\alpha}}
\def\Ub{\pmb{$U$}_{\beta}}
\def\uap{\pmb{$u$}^{\prime}_{\alpha}}
\def\Aa{A_{\alpha}}
\def\Ab{A_{\beta}}
\def\uabp{\pmb{$u$}^{\prime}_{\alpha \beta}}
\def\pabp{p^{\prime}_{\alpha \beta}}
\def\Ba{B_{\alpha}}
\def\Bas{B^*_{\alpha}}
\def\Ca{C_{\alpha}}
\def\Cas{C^*_{\alpha}}

\def\htf{\hskip .25truein}
\def\hf{\hskip .50truein}
\def\hsf{\hskip .75truein}

\def\inches{$^{\prime\prime}$}

\def\intlzt{\intl_0^t}
\def\intlv{\intl_V}
\def\ointlpv{\ointl_{\partial V}}
\def\ointlpb{\ointl_{\partial B}}
\def\sk{\pmb{{\cmmi k}}}
\def\sx{\pmb{{\cmmi x}}}

\newcommand{\dsty}{\displaystyle}
\newcommand{\bsub}{\begin{subequations}}
\newcommand{\esub}{\end{subequations}}

\newcommand{\al}{\alpha}
\newcommand{\be}{\beta}
\newcommand{\ka}{\kappa}
\newcommand{\om}{\omega}
\newcommand{\ga}{\gamma}
\newcommand{\de}{\delta}
\newcommand{\ep}{\epsilon}
\newcommand{\la}{\lambda}

\newcommand{\Ga}{\Gamma}

\newcommand{\po}{\mbox{\boldmath $\omega$}}
\newcommand{\peta}{\mbox{\boldmath $\eta$}}
\newcommand{\pzeta}{\mbox{\boldmath $\zeta$}}
\newcommand{\pt}{\mbox{\boldmath $\tau$}}
\newcommand{\pxi}{\mbox{\boldmath $\xi$}}
\newcommand{\ptau}{\mbox{\boldmath $\tau$}}
\newcommand{\ppsi}{\mbox{\boldmath $\psi$}}
\newcommand{\ps}{\mbox{\boldmath $\sigma$}}
\newcommand{\pg}{\mbox{\boldmath $\gamma$}}

\newcommand{\pGamma}{\mathbf \Gamma}

\newcommand{\bPhi}{\mathbf \Phi}
\newcommand{\pO}{\mathbf \Omega}

\newcommand{\bA}{\mathbf  A}
\newcommand{\bB}{\mathbf  B}
\newcommand{\bC}{\mathbf  C}
\newcommand{\bD}{\mathbf  D}
\newcommand{\bE}{\mathbf  E}
\newcommand{\bF}{\mathbf  F}
\newcommand{\bG}{\mathbf  G}
\newcommand{\bH}{\mathbf  H}
\newcommand{\bI}{\mathbf  I}
\newcommand{\bJ}{\mathbf  J}
\newcommand{\bK}{\mathbf  K}
\newcommand{\bL}{\mathbf  L}
\newcommand{\bP}{\mathbf  P}
\newcommand{\bQ}{\mathbf  Q}
\newcommand{\bR}{\mathbf  R}
\newcommand{\bS}{\mathbf  S}
\newcommand{\bT}{\mathbf  T}
\newcommand{\bV}{\mathbf  V}

\newcommand{\br}{\bar{r}  }
\newcommand{\bs}{\mathbf  s}

\newcommand{\cA}{\mathcal A}
\newcommand{\cB}{\mathcal B}
\newcommand{\cC}{\mathcal C}
\newcommand{\cD}{\mathcal D}
\newcommand{\cF}{\mathcal F}
\newcommand{\cH}{\mathcal H}
\newcommand{\cL}{\mathcal L}
\newcommand{\cM}{\mathcal M}
\newcommand{\cN}{\mathcal N}
\newcommand{\cV}{\mathcal V}
\newcommand{\cS}{\mathcal S}
\newcommand{\cW}{\mathcal W}

\newcommand{\pA}{\textbf{\emph{A}}}
\newcommand{\pB}{\textbf{\emph{B}}}
\newcommand{\pC}{\textbf{\emph{C}}}
\newcommand{\pE}{\textbf{\emph{E}}}
\newcommand{\pF}{\textbf{\emph{F}}}
\newcommand{\pG}{\textbf{\emph{G}}}
\newcommand{\pH}{\textbf{\emph{H}}}
\newcommand{\pI}{\textbf{\emph{I}}}
\newcommand{\pJ}{\textbf{\emph{J}}}
\newcommand{\pK}{\textbf{\emph{K}}}
\newcommand{\pL}{\textbf{\emph{L}}}
\newcommand{\pM}{\textbf{\emph{M}}}
\newcommand{\pQ}{\textbf{\emph{Q}}}
\newcommand{\pS}{\textbf{\emph{S}}}
\newcommand{\pX}{\textbf{\emph{X}}}
\newcommand{\pW}{\textbf{\emph{W}}}
\newcommand{\pU}{\textbf{\emph{U}}}
\newcommand{\pV}{\textbf{\emph{V}}}

\newcommand{\pa}{\textbf{\emph{a}}}
\newcommand{\pb}{\textbf{\emph{b}}}
\newcommand{\pc}{\textbf{\emph{c}}}
\newcommand{\pd}{\textbf{\emph{d}}}
\newcommand{\pe}{\textbf{\emph{e}}}
\newcommand{\pf}{\textbf{\emph{f}}}
\newcommand{\pgg}{\textbf{\emph{g}}}
\newcommand{\pph}{\textbf{\emph{h}}}
\newcommand{\pppi}{\textbf{\emph{i}}}
\newcommand{\pj}{\textbf{\emph{j}}}
\newcommand{\pk}{\textbf{\emph{k}}}
\newcommand{\hpk}{\hat{\textbf{\emph{k}}}}

\newcommand{\ppm}{\textbf{\emph{m}}}
\newcommand{\pn}{\textbf{\emph{n}}}
\newcommand{\pp}{\textbf{\emph{p}}}
\newcommand{\pq}{\textbf{\emph{q}}}
\newcommand{\pr}{\textbf{\emph{r}}}
\newcommand{\pss}{\textbf{\emph{s}}}
\newcommand{\pu}{\textbf{\emph{u}}}
\newcommand{\pv}{\textbf{\emph{v}}}
\newcommand{\pw}{\textbf{\emph{w}}}
\newcommand{\px}{\textbf{\emph{x}}}
\newcommand{\py}{\textbf{\emph{y}}}

\newcommand{\ptt}{\textbf{\emph{t}}}
\newcommand{\pbb}{\textbf{\emph{b}}}

\newcommand{\whn}{\hat{\textbf{\emph{n}}}}
\newcommand{\whh}{\hat{h}}
\newcommand{\whw}{\hat{w}}
\newcommand{\whal}{\hat{\alpha}}
\newcommand{\whga}{\hat{\gamma}}

\newcommand{\whC}{\hat{C}}
\newcommand{\whG}{\hat{G}}
\newcommand{\whGa}{\hat{\Gamma}}

\newcommand{\oom}{\overline{\om}}
\newcommand{\op}{\overline{p}}
\newcommand{\opu}{\overline{\textbf{\emph{u}}}}
\newcommand{\ou}{\overline{u}}
\newcommand{\oz}{\overline{z}} 
\newcommand{\ov}{\overline{v}} \newcommand{\oU}{\overline{U}}
\newcommand{\oV}{\overline{V}} \newcommand{\oq}{\overline{q}}
\newcommand{\oal}{\overline{\al}}
\newcommand{\obe}{\overline{\be}}
\newcommand{\ozeta}{\overline{\zeta}}
\newcommand{\jpu}{[\![\pu]\!]} \newcommand{\ju}{[\![u]\!]} \newcommand{\jv}{[\![v]\!]}
\newcommand{\jp}{[\![p]\!]} \newcommand{\jq}{[\![q]\!]}
\newcommand{\jphi}{[\![\phi]\!]}

\newcommand{\tdh}{\tilde{h}} \newcommand{\tdp}{\tilde{p}} \newcommand{\tds}{\tilde{s}}
\newcommand{\tdT}{\tilde{T}} \newcommand{\tdmu}{\tilde{\mu}}
\newcommand{\tdnu}{\tilde{\nu}} \newcommand{\tdrho}{\tilde{\rho}}

\newcommand{\pat}{\partial}
\newcommand{\na}{\nabla}
\newcommand{\x}{\times}
\newcommand{\cd}{\cdot}

\newcommand{\beq}{\begin{equation}}
\newcommand{\eeq}{\end{equation}}
\newcommand{\bsubeq}{\begin{subequations}}
\newcommand{\esubeq}{\end{subequations}}
\newcommand{\beqn}{\begin{eqnarray}}
\newcommand{\eeqn}{\end{eqnarray}}
\newcommand{\fr}{\frac}
\newcommand{\lb}{\label}
\newcommand{\er}{\eqref}


\preprint{APS/123-QED}

\title{Transition in Hypersonic Boundary Layers:\ Role of Dilatational Waves}
\affiliation{State Key Laboratory of Turbulence and Complex Systems,
Collaborative Innovation Center of Advanced Aero-Engine, Peking University, Beijing 100871, China}
\affiliation{Department of Mechanical Engineering, Virginia Commonwealth University}
\author{Chuanhong Zhang}
\altaffiliation[]{These authors contributed equally to this work.}
\author{Yiding Zhu}
\altaffiliation[]{These authors contributed equally to this work.}
\author{Huijing Yuan}
\author{Jiezhi Wu}
\author{Shiyi Chen}
\author{Cunbiao Lee}
\altaffiliation[ ]{cblee@mech.pku.edu.cn}
\affiliation{State Key Laboratory of Turbulence and Complex Systems,
Collaborative Innovation Center of Advanced Aero-Engine,\\Peking University, Beijing 100871, China}
\author{Mohamed Gad-el-Hak} \affiliation{Department of Mechanical \& Nuclear Engineering, Virginia Commonwealth University, Richmond, VA 23284, USA}
\date{\today}

\begin{abstract}
Transition and turbulence production in a hypersonic boundary layer is investigated in a Mach 6 quiet wind tunnel using Rayleigh-scattering visualization, fast-response pressure measurements, and particle image velocimetry.  It is found that the second instability acoustic mode is the key modulator of the transition process.  The second mode experiences a rapid growth and a very fast annihilation due to the effect of bulk viscosity. The second mode interacts strongly with the first vorticity mode to directly promote a fast growth of the latter and leads to immediate transition to turbulence.

\begin{description}

\item[PACS numbers]
 47.20.Ft, 47.40.Ki

\end{description}
\end{abstract}

\pacs{Valid PACS appear here}
\maketitle


Transition and turbulence production in hypersonic boundary layers have recently received considerable attention owing to their fundamental importance and strong relevance to the safety of  hypersonic vehicle flight, including significant increases in aerodynamic heating, entropy production, and drag \cite{Zhong2012, Fedorov2011}. Compared to incompressible flows \cite{Kachanov1994, Lee2008}, hypersonic transition is less understood due to additional complexities, such as second and higher instability modes, and nonlinear coupling of different processes. The problem presents  tremendous challenges to theory, experiment, and computation \cite{Maslov2001,Fedorov2003,Ma2005}.

During laminar-to-turbulence transition in compressible boundary layers, a second instability mode appears at very high frequency and becomes stronger with increasing Mach number. Unlike the conventional first-mode instability associated with transverse vortical waves, the second-mode instabilities are acoustic, longitudinal waves reflected between the relative sonic line and the solid wall,  and are of crucial importance at high Mach numbers \cite{Mack1969, Fedorov2011}. Detailed work comparing the linear growth of the second-mode with linear stability theory (LST) has been undertaken both in quiet \cite{Schneider2001, Schneider2013, Wilkinson1997, Alba2010} and noisy tunnels \cite{Zhang2013,Zhu2013,Stetson1992}.

Unfortunately, there exists conflicting understanding of the effects of the second-mode instability. Stetson \cite{Stetson1992} and Pruett $et$ $al$.\ \cite{Pruett1995} believed that the second mode instability plays a dominant role in transition, while Li \cite{Li2010}, Bountin \cite{Bountin2001}, and Dong $et$ $al$.\ \cite{Dong2007} took opposite view. Evidently, the dynamic process of the evolutions of the two modes and their interactions need to be clarified. Emanuel \cite{Emanuel1992} and Gad-el-Hak \cite{Gad-el-Hakl995}, among others, recognized the important role of the second coefficient of viscosity in high-Mach-number flows \cite{Rah1999}. By including a non-zero bulk viscosity as a correction term in a general, non-similar formulation of hypersonic laminar boundary layers, Emanuel \cite{Emanuel1992} found a heat transfer rate well in excess of that based on Stokes's hypothesis \cite{Stokes1845} prediction, i.e.\ assuming zero bulk viscosity. Despite these progresses, neither qualitative visualization nor quantitative measurement of the complete transition process has yet been reported, and an underlying mechanism of transition and its association to the bulk viscosity are still missing.

In this Letter, we report an experimental study of the second-mode instability and its relevance to turbulence production. We use combined visualization--measurement techniques to obtain the most complete and detailed data set.  We are able to display quantitatively the entire streamwise evolution of the first and second modes, from their appearance and growth to the transition to turbulence.  Through analysis, we reveal, for the first time, the underlying mechanism of the transition in hypersonic boundary layers and its strong connection to the second mode and the bulk viscosity.

\begin{figure}[b]
 \includegraphics[width=8cm]{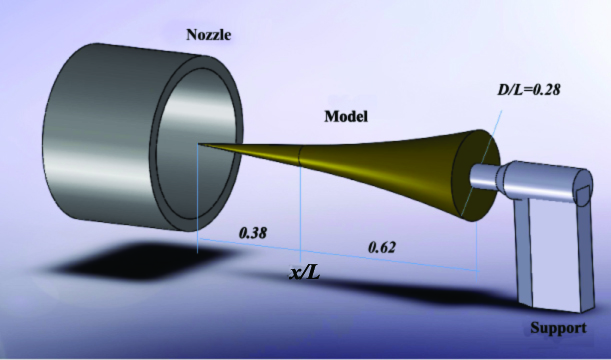}
\caption{\label{fig:cone} Schematic of the model. }
\end{figure}

\begin{figure*}
\includegraphics[width=17cm]{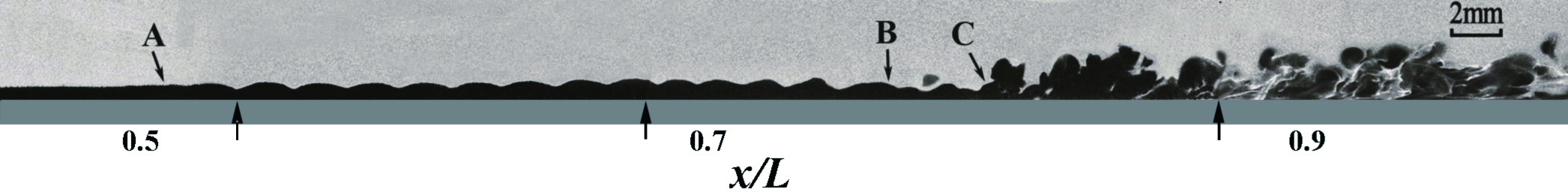}
\caption{\label{fig:visualization} Visualization of  transition stages. Arrow A:\ appearance of second mode; B:\ second mode's decay to almost zero; and C:\ onset of turbulence. The small vertical arrows indicate where a splice occurs.}
\end{figure*}

Our experimental model is a sharp-nosed flared cone, shown in Fig.~\ref{fig:cone}. Owing to the low freestream disturbance in the quiet flow, natural transition does not occur up to the end of an equivalent non-flared cone. Therefore, to obtain a whole picture of the transition process, a flared, instability-enhancing cone was studied under natural freestream disturbance environment. The freestream Mach number 6. The angle of attack is zero. The measurements were carried out in a newly established Mach 6 quiet wind tunnel (M6QT) at Peking University, which, at present, is one of three operational hypersonic quiet wind tunnels in the world  \cite{Schneider2013}.  A detailed description of the M6QT has been given by Zhang $et$ $al$.\ \cite{Zhang2013}. The sharp-nosed flared cone has a straight  $5^{\circ}$  half-angle for the first $0.38L$ of axial length, with a tangent flared region of radius $3.58L$ for the remaining $0.62L$, designed to enhance the instability and thus favor transition.  The total length of the cone is $L = 260$~mm. The origin of the coordinate system is located at the cone's nose, with $x$ being the streamwise coordinate along the cone's surface, $y$ being the coordinate normal to the cone's surface, and $z$ the transverse coordinate normal to the $x$-$y$ plane. The freestream stagnation temperature and pressure are, respectively, 430~K and 0.9~MPa. The freestream velocity  and unit Reynolds number are, respectively,  $870$ m/s and $9.0\times10^{6}$ per meter.

\begin{figure}[b]
\includegraphics[width=8cm]{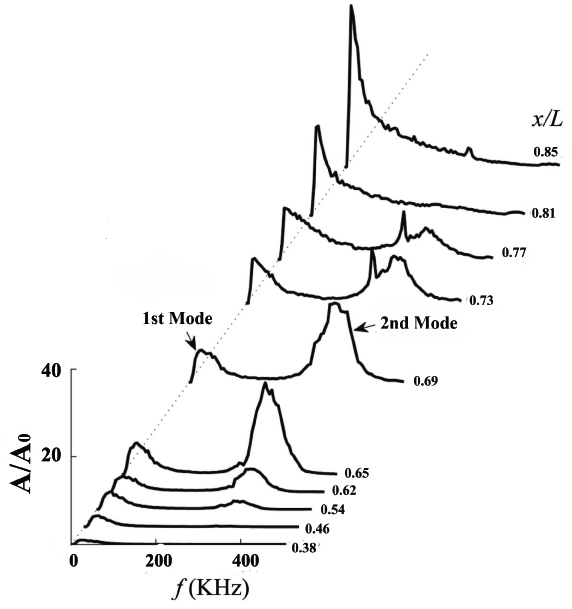}
\caption{\label{fig:PCB}  Frequency spectrum at different $x$-positions. Both first mode and second mode are shown. }
\end{figure}

In the first measurement reported here the whole transition process is investigated using Rayleigh-scattering techniques, which can offer a clear but qualitative information on the evolution of the structures in the boundary layer \cite{Smits2000}. In the Rayleigh-scattering images (RSI) shown in Fig.~\ref{fig:visualization}, independent measurements over four successive sections are spliced together to exhibit the whole evolution of the second mode. The flow is from left to right. In the upstream region, the boundary layer is laminar. Different from incompressible flow \cite{Lee2008}, a linear instability of the hypersonic flow results in growing second-mode patterns.  The unstable waves are weak at the early stage of the destabilized laminar flow, but eventually the amplitude of these waves increases dramatically. The instability waves are formed and persist for a long distance as shown in the region between arrow A and arrow B. The regular rope-like structures are second-mode waves and their wavelength is about 2 mm. After further nonlinear development, the second mode decays to small value as demonstrated in the region between arrows B and C, which will be quantitatively demonstrated in Fig.~\ref{fig:PCB}. The picture to the right of arrow C shows that a turbulent boundary layer appears immediately after transition. The transition process is similar for different unit Reynolds numbers. Because the second-mode waves exist for a long distance, a clear and complete visualization of the transition process has not been reported before.

\begin{figure*}[]
\includegraphics[width=15cm]{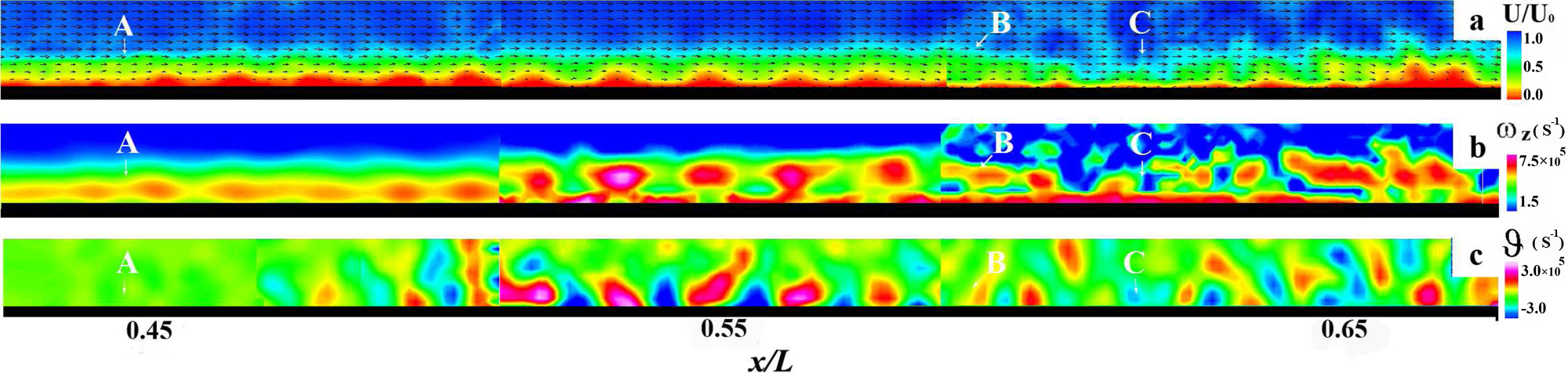}
\caption{\label{fig:PIV}(color online). PIV results of boundary layer development.  (a) Velocity magnitude normalized with freestream velocity; (b) spanwise component of vorticity $\omega_z$; (c) dilatation $\vartheta =\partial_x u+\partial_y v$. Arrow A:\ appearance of second mode; B:\ second mode's decay to almost zero; and C:\ onset of turbulence.}
\end{figure*}

\begin{figure*}[]
\includegraphics[width=15cm]{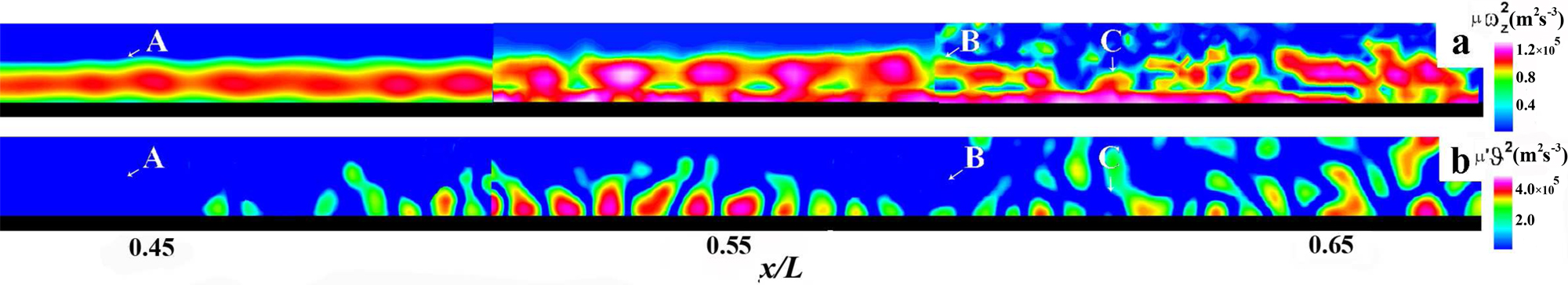}
\caption{\label{fig:PIV2}(color online). Viscosity dissipations induced by (a) vorticity; and (b) dilatation. Arrow A:\ appearance of second mode; B:\ second mode's decay to almost zero; and C:\ onset of turbulence.}
\end{figure*}

Surface-mounted PCB 132A31 fast-response piezoelectric pressure transducers were used to measure the pressure fluctuations on the model's surface. The first sensor is located at $x/L = 0.38$, and the first four are positioned 20~mm apart.  From $x/L = 0.62$, the spacing between the neighboring sensors is 10~mm to acquire a detailed evolution of the unstable disturbance. By analyzing the power spectral densities, we obtained detailed developments of frequencies and amplitudes of the first- and
second-mode disturbances, as partially shown in Fig.~\ref{fig:PCB}. Here the frequency results for the $x$-station locations down the model are shown with two distinct peaks, at 30~kHz and 350~kHz. The first and second modes are clearly shown. At a frequency of 30~kHz, the first mode grows continually to the onset of turbulence, while the second mode, at 350~kHz, grows initially and then decays very quickly, eventually reaching the  relatively quiet region.  The quick growth of the second mode is consistent with Mack's LST calculations \cite{Mack1984} and experimental results obtained using hot-wire anemometry \cite{Stetson1992}.

A second quantitative investigation of the transition process was carried out using a particle image velocimetry (PIV) system and PCB measurements simultaneously. The RSI and PCB results show that injection of tracer particles in PIV may cause the transition to occur earlier, but the whole transition process is almost the same as shown in Fig.~\ref{fig:visualization}. Based on this, a comparison between PCB data and PIV results can also be made. To solve the problem of very large in-plane displacement on PIV images, commonly encountered in hypersonic flows, we developed an improved image-preprocessing method \cite{Zhu2013}. Our strategy can extract the second-mode waves' fine flow structures from PIV raw images, while traditional PIV methods are incapable of doing that. PIV images were at first rotated horizontally and pre-processed by the new method. The interrogation started with 256$\times$256 pixels coarse samples, then 128$\times$128 pixels medium samples to capture the mean displacement fields from the time series. With the mean results as reference, instantaneous recordings were then evaluated with 64$\times$64 pixels samples to resolve the fine structures in the boundary layer. Smaller samples are unavailable because of the low particle density near the wall.  Instantaneous PIV results of the boundary layer development are shown in Fig.~\ref{fig:PIV}, where independent measurements over three successive sections are spliced together.

We plot in Fig.~\ref{fig:PIV} the velocity magnitude normalized by the freestream velocity, the $z$-component of vorticity, $\omega_z =\partial_x v -\partial_y u$, and dilatation $\vartheta =\partial_x u+\partial_y v$.  Note that if one writes $u=U(y)+u'$ with $U(y)$ being the mean shear velocity, then there is $\omega_z = -\partial_y U(y)+\omega'_z$, where $\omega'_z$ is the disturbance vorticity wave that is not shown in Fig.~\ref{fig:PIV}b. Combined with Figs.~\ref{fig:visualization}  and~\ref{fig:PCB}, these figures provide a clear evolution process of the first and second modes.  Notice a relatively quiet zone between arrows B and C with $x/L \in (0.60,0.62)$ in Figs.~\ref{fig:visualization} and \ref{fig:PIV}, where the boundary layer thickness becomes thin with the decay of the second mode. At the same time, the viscous dissipation is at a low level (see Fig.~\ref{fig:PIV2}).

Of the new phenomena observed in our experiments, the most remarkable ones are the rapid annihilation of the second (dilatation) mode after its fast growth, and the associated very strong enhancement of the first (vortical) mode that leads to a quick transition. Although a few experiments have shown the dominance of the second-mode waves in the production of hypersonic turbulence and their disappearance in turbulent boundary layers \cite{Alba2010}, the detailed evolution process of the second mode, especially its fast decay, has remained unclear. Schlieren or shadowgraph techniques can be used to observe the second mode \cite{Zhang2013}, but the fine flow structures in the boundary layer cannot be fully identified. Therefore, the above observed peculiar phenomena demand a physical interpretation, which we provide below.

In hypersonic boundary layers the second mode dominates the transitional flow, see Fig.~\ref{fig:PCB}. When the dilatation wave becomes sufficiently strong, it contributes considerably to the irreversible energy dissipation (Fig.\ref{fig:PIV2}). Indeed, it is well-known that, in addition to shear viscosity $\mu$, the compressibility of a viscous flow with nonzero dilatation brings in a longitudinal viscosity $\mu^{\prime}=\mu_{b}+\frac{4}{3} \mu$, where $\mu_b$ is the bulk viscosity, so that the pressure $p$ is modified to $p-\mu^{\prime}\vartheta$. Accordingly, the dissipation function $\Phi$ per unit volume can be explicitly split to the contributions of vorticity and dilatation \cite{wu2006, Lele1994}, where a divergence term of undetermined sign is neglected:
\begin{equation}
\Phi=\mu\omega^{2} +\mu^{\prime}\vartheta^{2} \ \ ;  \ \    \quad   \omega=|\bm \omega|,
\end{equation}
where $\mu$ is the shear viscosity. In monatomic gases, Stokes's hypothesis $\mu_b=0$ holds, implying $\mu^{\prime}=\frac{4}{3}\mu$. But in diatomic and polyatomic gases $\mu_{b}\neq0$. Both $\mu$ and $\mu_b$ are temperature dependent. Figure~\ref{fig:PIV2}a shows the shear-induced dissipation $\mu\omega^{2}$, which is largely independent of the compressibility but affected by temperature. The term $\mu^{\prime}\vartheta^{2}$ shown in
Fig.~\ref{fig:PIV2}b is a pure compressibility effect and represents the dilatation-induced dissipation. Here, a remarkable fact is that the bulk viscosity is not only a parameter of material property but also a dispersive function depending on frequency. While in a non-moving gas $\mu_{b} = O(\mu)$, it can be magnified significantly at high-frequency circumstances \cite{Landau1959}. The aforementioned high-frequency trapped acoustic waves relevant to hypersonic second-mode instability just provide such a circumstance. In our study, $\mu_{b}$  was chosen as $0.73\mu$ \cite{Cramer2012}, its value in quiet $N_{2}$ at $T=293$ K. Given the high frequency of 350~kHz and higher temperature, the value we use is certainly an over-conservative estimate and very likely to be much larger in hypersonic boundary layers (could be by a factor of $10^3$ higher).

\begin{figure}[t]
\includegraphics[width=9 cm]{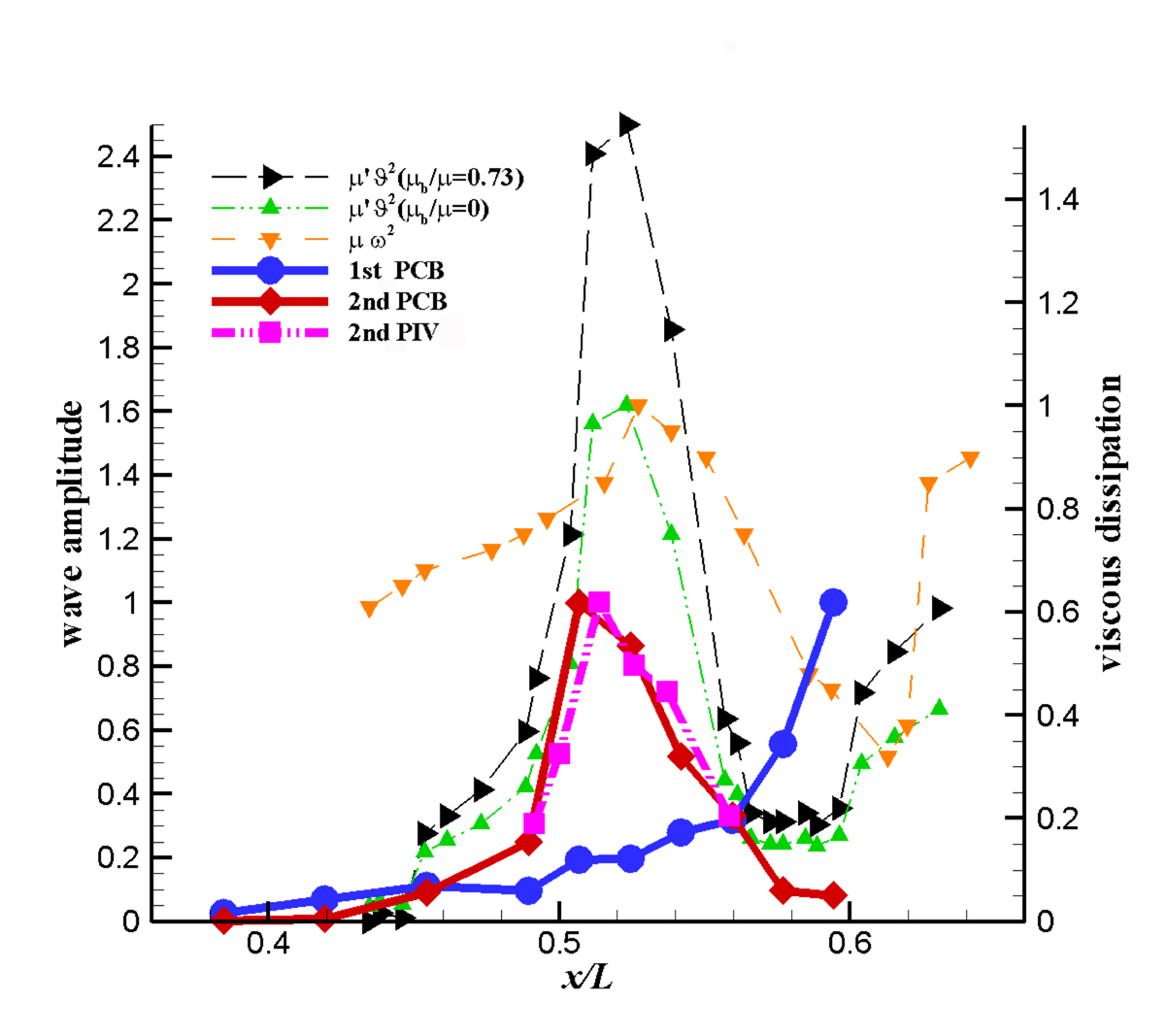}
\caption{\label{fig:PCBPIV}(color online). Comparison of wave amplitude and dissipation function by shear and dilatation.  Amplitude of first mode is taken from PCB, while that of second mode from both PCB and PIV.  Dissipation caused by dilatation is calculated for both $\mu_b =0.73\mu$ (an over-conservative estimate) and $\mu_b=0$.}
\end{figure}

Now, the amplitude of the instability waves and the viscous dissipation can be calculated by combining the simultaneous pressure and PIV measurements. Figure~\ref{fig:PCBPIV} displays more clearly the peak values of $\mu \omega^2$ and $\mu'\vartheta^2$ as well as their correlation with the spatial evolutions of first-  and second-instability-wave amplitudes.  Figure~\ref{fig:PIV2} shows the viscosity dissipation induced by vorticity (Fig.~\ref{fig:PIV2}a) and dissipation (Fig.~\ref{fig:PIV2}b).
The viscous dissipation values were extracted from Figs.~\ref{fig:PIV2}a and~\ref{fig:PIV2}b. The dissipation caused by $\mu'\vartheta^2$ was sampled at the local peak points of the region where $\vartheta\neq 0$. Each quantity was normalized by its maximum value.

Physically, the aforementioned modal interactions can be explained by the strong $(\bm\omega,\vartheta)$-couplings that exist generically both in the interior of the boundary layer and at the wall \cite{wu2006}. The internal coupling occurs via the compressible vorticity transport equation.  In the flow region away from the viscous sublayer with negligible normal motion, the disturbance vorticity $\omega'_z$ is governed by

\begin{equation}
\frac{\partial \omega'_z}{\partial t}+ U(y)\frac{\partial \omega'_z}{\partial x}+\vartheta\omega'_z \simeq 0
\label{eq:vt}
\end{equation}
Our experiment has found that in the range of  $x/L  \in (0.47,0.53)$, the vortical and acoustic waves have almost the same phase speed $c$ \cite{Zhang2014}, so both can be approximately expressed as functions of $\eta = x-ct$, with $c\neq u$ in this region. Thus, (\ref{eq:vt}) yields the following coupling relation between the phase-locked traveling $\omega_z$-wave and $\vartheta$-wave

\begin{equation}\label{eq:u-c}
\omega'_z(\eta) \propto \exp\left(-\int^\eta_0 \frac{\vartheta(\zeta) d\zeta}{U-c}\right)
\end{equation}
indicating that the $\omega$-wave can be significantly magnified by the $\vartheta$-wave at its peak-value zone and near the critical layer. The theory of the phase-locked nonlinear interaction has been well developed, known as nonlinear critical-layer analysis (see the review by \cite{wu2004}). It states that the second mode will significantly prompt the growth of the first mode. Our experimental results show that the small-amplitude first mode has gone through a remarkable growth while the large-amplitude second mode decays sharply, indicating that the energy has been transferred from the high-frequency disturbances to the low-frequency disturbances. This can be viewed as a direct validation of the theory.

On the other hand, the boundary $(\om,p)$-coupling occurs via viscosity and the no-slip condition, which can be estimated more accurately than Eq.~(3).
Consider the interaction of two traveling waves of the form of $\eta =x-ct$\/: the normal-stress wave $\Pi\equiv p-\mu'\vartheta =\Pi(k_\theta\eta)$ and the shear-stress wave $\nu\om(k_\om \eta)$, with $k_\theta\gg 1$ and $k_\om\ll k_\theta$ being the wave numbers of second and first modes, respectively. The vorticity creation rate at the wall is measured by $\sigma \equiv -\nu\pat_y \om$ at $y=0$[34]. Then due to the no-slip condition, the on-wall viscous $(\om,p)$-coupling takes the form[29,34]
\beq\lb{C-Ra}
\sigma(x,t) =-\nu\fr{\pat \om}{\pat y} = \fr{\pat \Pi}{\pat x} = k_\theta\Pi'(\eta)\quad {\rm at}\ y=0,
\eeq
indicating that the high-frequency $\vartheta$-wave must create new $\om$-wave of the same $k_\theta$, resulting in a high-frequency Stokes layer, which directly prepares the appearance of strong turbulent fluctuations. The extra large bulk viscosity $\mu_b$ must significantly enhance this boundary coupling.

With these
$(\om,p)$-couplings inside the flow and at the wall, why the
first-mode instability with small vorticity wave number $k_\om\ll k_\theta$ is quickly enhanced and gains high-frequency components right after the
dissipation of second mode can now be understood. It may be mentioned that symmetric to Eq.~(1) there is
\beq\lb{C-Rb}
\fr{\pat \Pi}{\pat y} = \nu k_\om \om'(\eta)\quad {\rm at}\ y=0,
\eeq
which represents how the low-frequency first-mode wave alters the Neumann condition of the dilatation equation. But since $\nu k_\om \sim Re^{-1/2}k_\om \ll 1$, this effect is pretty weak or even negligible. This confirms the preceding analysis on the one-way interaction of fast-mode and slow-mode instability waves.

In summary, our results suggest the following route from second-mode instability to turbulence production. At first, the free-stream disturbances penetrate into
the boundary layer near the leading edge of the cone and then evolves as the linear stability theory predicts. The unstable waves (the first mode and the
second mode) will be amplified. When the amplitudes of these unstable become large enough, the nonlinear interactions may be triggered. The nonlinear coupling of dilatation waves and vortical waves inside the boundary layer and their linear coupling at the wall enable the highly localized strong peak of dilatation wave to enhance the first-mode instability and thereby cause the production of turbulence. In particular, the bulk viscosity $\mu_{b}$, which has so far been overlooked by most theoretical studies and numerical simulations, serves as a strong magnifier not only in the very strong dissipation peak of high-frequency dilatation waves but also in the viscous $(\bm \omega, \vartheta)$-coupling at the wall. The proposed scenario may inspire new strategies for the control of transition in hypersonic boundary layers.

\begin{acknowledgments}
The authors wish to thank Profs. W.\ D.\ Su and J.\ Su for their valuable suggestions and very fruitful discussions. This work was supported by the National Natural Science Foundation of China under grant no.\ 109103010062, and by the National Climb-B Plan under grant no.\ 2009CB724100.
\end{acknowledgments}

\nocite{*}

\bibliography{apssamp}

\providecommand{\noopsort}[1]{}\providecommand{\singleletter}[1]{#1}%
\begin{thebibliography}{32}%
\makeatletter
\providecommand \@ifxundefined [1]{%
 \@ifx{#1\undefined}
}%
\providecommand \@ifnum [1]{%
 \ifnum #1\expandafter \@firstoftwo
 \else \expandafter \@secondoftwo
 \fi
}%
\providecommand \@ifx [1]{%
 \ifx #1\expandafter \@firstoftwo
 \else \expandafter \@secondoftwo
 \fi
}%
\providecommand \natexlab [1]{#1}%
\providecommand \enquote  [1]{``#1''}%
\providecommand \bibnamefont  [1]{#1}%
\providecommand \bibfnamefont [1]{#1}%
\providecommand \citenamefont [1]{#1}%
\providecommand \href@noop [0]{\@secondoftwo}%
\providecommand \href [0]{\begingroup \@sanitize@url \@href}%
\providecommand \@href[1]{\@@startlink{#1}\@@href}%
\providecommand \@@href[1]{\endgroup#1\@@endlink}%
\providecommand \@sanitize@url [0]{\catcode `\\12\catcode `\$12\catcode
  `\&12\catcode `\#12\catcode `\^12\catcode `\_12\catcode `\%12\relax}%
\providecommand \@@startlink[1]{}%
\providecommand \@@endlink[0]{}%
\providecommand \url  [0]{\begingroup\@sanitize@url \@url }%
\providecommand \@url [1]{\endgroup\@href {#1}{\urlprefix }}%
\providecommand \urlprefix  [0]{URL }%
\providecommand \Eprint [0]{\href }%
\providecommand \doibase [0]{http://dx.doi.org/}%
\providecommand \selectlanguage [0]{\@gobble}%
\providecommand \bibinfo  [0]{\@secondoftwo}%
\providecommand \bibfield  [0]{\@secondoftwo}%
\providecommand \translation [1]{[#1]}%
\providecommand \BibitemOpen [0]{}%
\providecommand \bibitemStop [0]{}%
\providecommand \bibitemNoStop [0]{.\EOS\space}%
\providecommand \EOS [0]{\spacefactor3000\relax}%
\providecommand \BibitemShut  [1]{\csname bibitem#1\endcsname}%
\let\auto@bib@innerbib\@empty
\bibitem [{\citenamefont {Zhong}\ and\ \citenamefont {Wang}(2012)}]{Zhong2012}%
  \BibitemOpen
  \bibfield  {author} {\bibinfo {author} {\bibfnamefont {X.}~\bibnamefont
  {Zhong}}\ and\ \bibinfo {author} {\bibfnamefont {X.}~\bibnamefont {Wang}},\
  }\href@noop {} {\bibfield  {journal} {\bibinfo  {journal} {Annu. Rev. Fluid
  Mech.}\ }\textbf {\bibinfo {volume} {44}},\ \bibinfo {pages} {527} (\bibinfo
  {year} {2012})}\BibitemShut {NoStop}%
\bibitem [{\citenamefont {Fedorov}(2011)}]{Fedorov2011}%
  \BibitemOpen
  \bibfield  {author} {\bibinfo {author} {\bibfnamefont {A.~V.}\ \bibnamefont
  {Fedorov}},\ }\href@noop {} {\bibfield  {journal} {\bibinfo  {journal} {Annu.
  Rev. Fluid Mech.}\ }\textbf {\bibinfo {volume} {43}},\ \bibinfo {pages} {79}
  (\bibinfo {year} {2011})}\BibitemShut {NoStop}%
\bibitem [{\citenamefont {Kachanov}(1994)}]{Kachanov1994}%
  \BibitemOpen
  \bibfield  {author} {\bibinfo {author} {\bibfnamefont {Y.~S.}\ \bibnamefont
  {Kachanov}},\ }\href@noop {} {\bibfield  {journal} {\bibinfo  {journal}
  {Annu. Rev. Fluid Mech.}\ }\textbf {\bibinfo {volume} {26}},\ \bibinfo
  {pages} {411} (\bibinfo {year} {1994})}\BibitemShut {NoStop}%
\bibitem [{\citenamefont {Lee}\ and\ \citenamefont {Wu}(2008)}]{Lee2008}%
  \BibitemOpen
  \bibfield  {author} {\bibinfo {author} {\bibfnamefont {C.~B.}\ \bibnamefont
  {Lee}}\ and\ \bibinfo {author} {\bibfnamefont {J.~Z.}\ \bibnamefont {Wu}},\
  }\href@noop {} {\bibfield  {journal} {\bibinfo  {journal} {Appl. Mech. Rev.}\
  }\textbf {\bibinfo {volume} {61}},\ \bibinfo {pages} {030802} (\bibinfo
  {year} {2008})}\BibitemShut {NoStop}%
\bibitem [{\citenamefont {Maslov}\ \emph {et~al.}(2001)\citenamefont {Maslov},
  \citenamefont {Shiplyuk}, \citenamefont {Sidorenko},\ and\ \citenamefont
  {Arnal}}]{Maslov2001}%
  \BibitemOpen
  \bibfield  {author} {\bibinfo {author} {\bibfnamefont {A.~A.}\ \bibnamefont
  {Maslov}}, \bibinfo {author} {\bibfnamefont {A.~N.}\ \bibnamefont
  {Shiplyuk}}, \bibinfo {author} {\bibfnamefont {A.~A.}\ \bibnamefont
  {Sidorenko}}, \ and\ \bibinfo {author} {\bibfnamefont {D.}~\bibnamefont
  {Arnal}},\ }\href@noop {} {\bibfield  {journal} {\bibinfo  {journal} {J.
  Fluid Mech.}\ }\textbf {\bibinfo {volume} {426}},\ \bibinfo {pages} {73}
  (\bibinfo {year} {2001})}\BibitemShut {NoStop}%
\bibitem [{\citenamefont {Fedorov}(2003)}]{Fedorov2003}%
  \BibitemOpen
  \bibfield  {author} {\bibinfo {author} {\bibfnamefont {A.~V.}\ \bibnamefont
  {Fedorov}},\ }\href@noop {} {\bibfield  {journal} {\bibinfo  {journal} {J.
  Fluid Mech.}\ }\textbf {\bibinfo {volume} {491}},\ \bibinfo {pages} {101}
  (\bibinfo {year} {2003})}\BibitemShut {NoStop}%
\bibitem [{\citenamefont {Ma}\ and\ \citenamefont {Zhong}(2005)}]{Ma2005}%
  \BibitemOpen
  \bibfield  {author} {\bibinfo {author} {\bibfnamefont {Y.}~\bibnamefont
  {Ma}}\ and\ \bibinfo {author} {\bibfnamefont {X.}~\bibnamefont {Zhong}},\
  }\href@noop {} {\bibfield  {journal} {\bibinfo  {journal} {J. Fluid Mech.}\
  }\textbf {\bibinfo {volume} {532}},\ \bibinfo {pages} {63} (\bibinfo {year}
  {2005})}\BibitemShut {NoStop}%
\bibitem [{\citenamefont {Mack}(1969)}]{Mack1969}%
  \BibitemOpen
  \bibfield  {author} {\bibinfo {author} {\bibfnamefont {L.~M.}\ \bibnamefont
  {Mack}},\ }\href@noop {} {}\bibinfo {type} {{Jet Propul. Lab. Doc.}}\
  \bibinfo {number} {900-277}\ (\bibinfo {year} {1969})\BibitemShut {NoStop}%
\bibitem [{\citenamefont {Schneider}(2001)}]{Schneider2001}%
  \BibitemOpen
  \bibfield  {author} {\bibinfo {author} {\bibfnamefont {S.~P.}\ \bibnamefont
  {Schneider}},\ }\href@noop {} {\bibfield  {journal} {\bibinfo  {journal} {J.
  Spacecr. Rockets}\ }\textbf {\bibinfo {volume} {38(3)}},\ \bibinfo {pages}
  {323} (\bibinfo {year} {2001})}\BibitemShut {NoStop}%
\bibitem [{\citenamefont {Schneider}(2013)}]{Schneider2013}%
  \BibitemOpen
  \bibfield  {author} {\bibinfo {author} {\bibfnamefont {S.~P.}\ \bibnamefont
  {Schneider}},\ }\href@noop {} {\bibfield  {journal} {\bibinfo  {journal}
  {AIAA Pap.}\ }\textbf {\bibinfo {volume} {2608}} (\bibinfo {year}
  {2013})}\BibitemShut {NoStop}%
\bibitem [{\citenamefont {Wilkinson}(1997)}]{Wilkinson1997}%
  \BibitemOpen
  \bibfield  {author} {\bibinfo {author} {\bibfnamefont {S.~P.}\ \bibnamefont
  {Wilkinson}},\ }\href@noop {} {\bibfield  {journal} {\bibinfo  {journal}
  {AIAA Pap.}\ }\textbf {\bibinfo {volume} {1819}} (\bibinfo {year}
  {1997})}\BibitemShut {NoStop}%
\bibitem [{\citenamefont {Alba}\ \emph {et~al.}(2010)\citenamefont {Alba},
  \citenamefont {Casper}, \citenamefont {Beresh},\ and\ \citenamefont
  {Schneider}}]{Alba2010}%
  \BibitemOpen
  \bibfield  {author} {\bibinfo {author} {\bibfnamefont {C.~R.}\ \bibnamefont
  {Alba}}, \bibinfo {author} {\bibfnamefont {K.~M.}\ \bibnamefont {Casper}},
  \bibinfo {author} {\bibfnamefont {S.~J.}\ \bibnamefont {Beresh}}, \ and\
  \bibinfo {author} {\bibfnamefont {S.~P.}\ \bibnamefont {Schneider}},\
  }\href@noop {} {\bibfield  {journal} {\bibinfo  {journal} {AIAA Pap.}\
  }\textbf {\bibinfo {volume} {897}} (\bibinfo {year} {2010})}\BibitemShut
  {NoStop}%
\bibitem [{\citenamefont {Zhang}\ \emph {et~al.}(2013)\citenamefont {Zhang},
  \citenamefont {Tang},\ and\ \citenamefont {Lee}}]{Zhang2013}%
  \BibitemOpen
  \bibfield  {author} {\bibinfo {author} {\bibfnamefont {C.~H.}\ \bibnamefont
  {Zhang}}, \bibinfo {author} {\bibfnamefont {Q.}~\bibnamefont {Tang}}, \ and\
  \bibinfo {author} {\bibfnamefont {C.~B.}\ \bibnamefont {Lee}},\ }\href@noop
  {} {\bibfield  {journal} {\bibinfo  {journal} {Acta Mech. Sin.}\ }\textbf
  {\bibinfo {volume} {29(1)}},\ \bibinfo {pages} {48} (\bibinfo {year}
  {2013})}\BibitemShut {NoStop}%
\bibitem [{\citenamefont {Zhu}\ \emph {et~al.}(2013)\citenamefont {Zhu},
  \citenamefont {Yuan}, \citenamefont {Zhang},\ and\ \citenamefont
  {Lee}}]{Zhu2013}%
  \BibitemOpen
  \bibfield  {author} {\bibinfo {author} {\bibfnamefont {Y.~D.}\ \bibnamefont
  {Zhu}}, \bibinfo {author} {\bibfnamefont {H.~J.}\ \bibnamefont {Yuan}},
  \bibinfo {author} {\bibfnamefont {C.~H.}\ \bibnamefont {Zhang}}, \ and\
  \bibinfo {author} {\bibfnamefont {C.~B.}\ \bibnamefont {Lee}},\ }\href@noop
  {} {\bibfield  {journal} {\bibinfo  {journal} {Meas. Sci. Technol.}\ }\textbf
  {\bibinfo {volume} {24}},\ \bibinfo {pages} {125302} (\bibinfo {year}
  {2013})}\BibitemShut {NoStop}%
\bibitem [{\citenamefont {Stetson}\ and\ \citenamefont
  {Kimmel}(1992)}]{Stetson1992}%
  \BibitemOpen
  \bibfield  {author} {\bibinfo {author} {\bibfnamefont {K.~F.}\ \bibnamefont
  {Stetson}}\ and\ \bibinfo {author} {\bibfnamefont {R.~L.}\ \bibnamefont
  {Kimmel}},\ }\href@noop {} {\bibfield  {journal} {\bibinfo  {journal} {AIAA
  Pap.}\ }\textbf {\bibinfo {volume} {0737}} (\bibinfo {year}
  {1992})}\BibitemShut {NoStop}%
\bibitem [{\citenamefont {Pruett}\ and\ \citenamefont
  {Chang}(1995)}]{Pruett1995}%
  \BibitemOpen
  \bibfield  {author} {\bibinfo {author} {\bibfnamefont {C.~D.}\ \bibnamefont
  {Pruett}}\ and\ \bibinfo {author} {\bibfnamefont {C.~L.}\ \bibnamefont
  {Chang}},\ }\href@noop {} {\bibfield  {journal} {\bibinfo  {journal}
  {Theoretical and Computational Fluid Dynamics.}\ }\textbf {\bibinfo {volume}
  {7(5)}},\ \bibinfo {pages} {397} (\bibinfo {year} {1995})}\BibitemShut
  {NoStop}%
\bibitem [{\citenamefont {Li}\ \emph {et~al.}(2010)\citenamefont {Li},
  \citenamefont {Fu},\ and\ \citenamefont {Ma}}]{Li2010}%
  \BibitemOpen
  \bibfield  {author} {\bibinfo {author} {\bibfnamefont {X.}~\bibnamefont
  {Li}}, \bibinfo {author} {\bibfnamefont {D.}~\bibnamefont {Fu}}, \ and\
  \bibinfo {author} {\bibfnamefont {Y.}~\bibnamefont {Ma}},\ }\href@noop {}
  {\bibfield  {journal} {\bibinfo  {journal} {Physics of Fluids}\ }\textbf
  {\bibinfo {volume} {22}},\ \bibinfo {pages} {025105} (\bibinfo {year}
  {2010})}\BibitemShut {NoStop}%
\bibitem [{\citenamefont {Bountin}\ \emph {et~al.}(2001)\citenamefont
  {Bountin}, \citenamefont {Sidorenko},\ and\ \citenamefont
  {Shiplyuk}}]{Bountin2001}%
  \BibitemOpen
  \bibfield  {author} {\bibinfo {author} {\bibfnamefont {D.~A.}\ \bibnamefont
  {Bountin}}, \bibinfo {author} {\bibfnamefont {A.~A.}\ \bibnamefont
  {Sidorenko}}, \ and\ \bibinfo {author} {\bibfnamefont {A.~N.}\ \bibnamefont
  {Shiplyuk}},\ }\href@noop {} {\bibfield  {journal} {\bibinfo  {journal}
  {Journal of Applied Mechanics and Technical Physics.}\ }\textbf {\bibinfo
  {volume} {42(1)}},\ \bibinfo {pages} {57�C62} (\bibinfo {year}
  {2001})}\BibitemShut {NoStop}%
\bibitem [{\citenamefont {D.~Ming}(2007)}]{Dong2007}%
  \BibitemOpen
  \bibfield  {author} {\bibinfo {author} {\bibfnamefont {J.~S.~L.}\
  \bibnamefont {D.~Ming}},\ }\href@noop {} {\bibfield  {journal} {\bibinfo
  {journal} {Applied Mathematics and Mechanics(English Edition).}\ }\textbf
  {\bibinfo {volume} {28(8)}},\ \bibinfo {pages} {1019�C1028} (\bibinfo
  {year} {2007})}\BibitemShut {NoStop}%
\bibitem [{\citenamefont {Emanuel}(1992)}]{Emanuel1992}%
  \BibitemOpen
  \bibfield  {author} {\bibinfo {author} {\bibfnamefont {G.}~\bibnamefont
  {Emanuel}},\ }\href@noop {} {\bibfield  {journal} {\bibinfo  {journal}
  {Physics of Fluids A}\ }\textbf {\bibinfo {volume} {4}},\ \bibinfo {pages}
  {491} (\bibinfo {year} {1992})}\BibitemShut {NoStop}%
\bibitem [{\citenamefont {Gad-el Hak}(1995)}]{Gad-el-Hakl995}%
  \BibitemOpen
  \bibfield  {author} {\bibinfo {author} {\bibfnamefont {M.}~\bibnamefont
  {Gad-el Hak}},\ }\href@noop {} {\bibfield  {journal} {\bibinfo  {journal}
  {Journal of Fluids Engineering}\ }\textbf {\bibinfo {volume} {117}},\
  \bibinfo {pages} {3} (\bibinfo {year} {1995})}\BibitemShut {NoStop}%
\bibitem [{\citenamefont {Rah}\ and\ \citenamefont {Eu}(1999)}]{Rah1999}%
  \BibitemOpen
  \bibfield  {author} {\bibinfo {author} {\bibfnamefont {K.}~\bibnamefont
  {Rah}}\ and\ \bibinfo {author} {\bibfnamefont {B.~C.}\ \bibnamefont {Eu}},\
  }\href@noop {} {\bibfield  {journal} {\bibinfo  {journal} {Phys. Rev. Lett.}\
  }\textbf {\bibinfo {volume} {83}},\ \bibinfo {pages} {4566} (\bibinfo {year}
  {1999})}\BibitemShut {NoStop}%
\bibitem [{\citenamefont {Stokes}(1845)}]{Stokes1845}%
  \BibitemOpen
  \bibfield  {author} {\bibinfo {author} {\bibfnamefont {G.~G.}\ \bibnamefont
  {Stokes}},\ }\href@noop {} {\bibfield  {journal} {\bibinfo  {journal} {Trans.
  Camb. Phil. Soc.}\ }\textbf {\bibinfo {volume} {8}},\ \bibinfo {pages} {287}
  (\bibinfo {year} {1845})}\BibitemShut {NoStop}%
\bibitem [{\citenamefont {Smits}\ and\ \citenamefont {Lim}(2000)}]{Smits2000}%
  \BibitemOpen
  \bibfield  {author} {\bibinfo {author} {\bibfnamefont {A.~J.}\ \bibnamefont
  {Smits}}\ and\ \bibinfo {author} {\bibfnamefont {T.~T.}\ \bibnamefont
  {Lim}},\ }\href@noop {} {\emph {\bibinfo {title} {Flow visualization:
  techniques and examples}}}\ (\bibinfo  {publisher} {Imperial College Press},\
  \bibinfo {year} {2000})\BibitemShut {NoStop}%
\bibitem [{\citenamefont {Mack}(1984)}]{Mack1984}%
  \BibitemOpen
  \bibfield  {author} {\bibinfo {author} {\bibfnamefont {L.~M.}\ \bibnamefont
  {Mack}},\ }\href@noop {} {\emph {\bibinfo {title} {Boundary-layer linear
  stability theory}}},\ \bibinfo {type} {{Rep. AGARD}}\ \bibinfo {number}
  {709}\ (\bibinfo  {institution} {Jet Propul. Lab., Pasadena, CA},\ \bibinfo
  {year} {1984})\BibitemShut {NoStop}%
\bibitem [{\citenamefont {Wu}\ \emph {et~al.}(2006)\citenamefont {Wu},
  \citenamefont {Ma},\ and\ \citenamefont {Zhou}}]{wu2006}%
  \BibitemOpen
  \bibfield  {author} {\bibinfo {author} {\bibfnamefont {J.~Z.}\ \bibnamefont
  {Wu}}, \bibinfo {author} {\bibfnamefont {H.~Y.}\ \bibnamefont {Ma}}, \ and\
  \bibinfo {author} {\bibfnamefont {M.~D.}\ \bibnamefont {Zhou}},\ }\href@noop
  {} {\emph {\bibinfo {title} {Vorticity and vortex dynamics}}}\ (\bibinfo
  {publisher} {Springer},\ \bibinfo {year} {2006})\BibitemShut {NoStop}%
\bibitem [{\citenamefont {Lele}(1994)}]{Lele1994}%
  \BibitemOpen
  \bibfield  {author} {\bibinfo {author} {\bibfnamefont {S.~K.}\ \bibnamefont
  {Lele}},\ }\href@noop {} {\bibfield  {journal} {\bibinfo  {journal} {Annu.
  Rev. Fluid Mech.}\ }\textbf {\bibinfo {volume} {26}},\ \bibinfo {pages} {211}
  (\bibinfo {year} {1994})}\BibitemShut {NoStop}%
\bibitem [{\citenamefont {Landau}\ and\ \citenamefont
  {Lifshitz}(1959)}]{Landau1959}%
  \BibitemOpen
  \bibfield  {author} {\bibinfo {author} {\bibfnamefont {L.~D.}\ \bibnamefont
  {Landau}}\ and\ \bibinfo {author} {\bibfnamefont {E.~M.}\ \bibnamefont
  {Lifshitz}},\ }\href@noop {} {\emph {\bibinfo {title} {Fluid Mechanics}}}\
  (\bibinfo  {publisher} {Pergamon},\ \bibinfo {year} {1959})\BibitemShut
  {NoStop}%
\bibitem [{\citenamefont {Cramer}(2012)}]{Cramer2012}%
  \BibitemOpen
  \bibfield  {author} {\bibinfo {author} {\bibfnamefont {M.~S.}\ \bibnamefont
  {Cramer}},\ }\href@noop {} {\bibfield  {journal} {\bibinfo  {journal} {Phys.
  Fluids}\ }\textbf {\bibinfo {volume} {24}},\ \bibinfo {pages} {066102}
  (\bibinfo {year} {2012})}\BibitemShut {NoStop}%
\bibitem [{\citenamefont {Zhang}(2014)}]{Zhang2014}%
  \BibitemOpen
  \bibfield  {author} {\bibinfo {author} {\bibfnamefont {C.}~\bibnamefont
  {Zhang}},\ }\emph {\bibinfo {title} {The development of hypersonic quiet wind
  tunnel and experimental investigation of hypersonic boundary-layer transition
  on flared cone}},\ \href@noop {} {\bibinfo {type} {{Ph.D.} thesis}},\
  \bibinfo  {school} {Peking University} (\bibinfo {year} {2014})\BibitemShut
  {NoStop}%
\bibitem [{\citenamefont {Wu}(2004)}]{wu2004}%
  \BibitemOpen
  \bibfield  {author} {\bibinfo {author} {\bibfnamefont {X.}~\bibnamefont
  {Wu}},\ }\href@noop {} {\bibfield  {journal} {\bibinfo  {journal} {Acta
  Mechanica Sinica}\ }\textbf {\bibinfo {volume} {20}},\ \bibinfo {pages} {327}
  (\bibinfo {year} {2004})}\BibitemShut {NoStop}%
\bibitem [{\citenamefont {Lighthill}(1963)}]{lighthill1963}%
  \BibitemOpen
  \bibfield  {author} {\bibinfo {author} {\bibfnamefont {M.~J.}\ \bibnamefont
  {Lighthill}},\ }\enquote {\bibinfo {title} {Introduction of boundary layer
  theory},}\ in\ \href@noop {} {\emph {\bibinfo {booktitle} {Laminar Boundary
  Layers}}},\ \bibinfo {editor} {edited by\ \bibinfo {editor} {\bibfnamefont
  {L.}~\bibnamefont {Rosenhead}}}\ (\bibinfo  {publisher} {Oxford University
  Press},\ \bibinfo {address} {New York},\ \bibinfo {year} {1963})\ p.~\bibinfo
  {pages} {46}\BibitemShut {NoStop}%
\end{thebibliography}%

\end{document}